\documentstyle[12pt]{l-aa}
\newcommand{\EQ}{\begin{equation}}
\newcommand{\EN}{\end{equation}}
\newcommand{\EQA}{\begin{eqnarray}}
\newcommand{\ENA}{\end{eqnarray}}

\input psfig
\begin{document}
\thesaurus{12.03.3, 12.07.1, 12.12.1}
\title{ Detection of Weak Lensing in the Fields of Luminous
Radiosources
}
\author{ B.\ Fort\inst{1}
\and  Y.\ Mellier\inst{2}
\and  M. \ Dantel-Fort\inst{1}
\and H.\ Bonnet\inst{2}
\and J.-P.\ Kneib\inst{3}}
\offprints{B. Fort}
\institute{Observatoire de Paris
DEMIRM.
61, avenue de l'Observatoire, F-75014 Paris, France
\and
  Observatoire Midi-Pyr\'en\'ees
LAT-URA285.
14, avenue Edouard Belin F-31400 Toulouse, France
\and
  Institute of Astronomy. Madingley Road,
Cambridge CB3 0HA, UK. }

\date{ Received ?? ; accepted ?? }
\maketitle
\footnote{Based on observations at the Canada-France-Hawaii
Telescope operated by the french Institut des Sciences de l'Univers
(INSU), the   Canadian National Research Council (CNRC)  and the
University of Hawaii (UH), and at the European Southern Observatory (ESO)
at La Silla-Chili.}
\begin{abstract}

We present a first attempt to reveal
the possible existence of large foreground mass condensations
 directly responsible for the  gravitational
magnification of four distant luminous radiosources and one optical QSO.
The technique uses a weak lensing analysis
of the distant galaxies in the field of each source.
We find a coherent shear
\footnote{Following the \cite*{bonnet94} and \cite*{bonnet95} papers,
 we will use the term
{\it shear} or {\it polarisation} as $e=1-b/a$.}
map with a large  magnification bias
on the line of sight to Q1622+0328. The local shear
in the field of the bright radiosources is also often
correlated with nearby groups or poor clusters of galaxies.
For three of them: PKS0135-247, 3C446 and PKS1508-05,
the groups are identified as gravitational deflectors
that magnify the radiosources. \par
This  result suggests that a substantial amount of invisible mass is
condensed in groups and poor clusters of galaxies. It may
explain the origin of a large angular correlation between
the distribution of distant radiosources ($z>$1) and
the distribution of low
redshift galaxies ($z<$0.3) (Bartelmann \& Schneider 1993).  \par
 We discuss the  feasability and consequences of a future systematic survey
to investigate the problem of magnification bias in the fields of
 luminous distant objects and to probe
the mass distributions  of galaxy groups at intermediate redshifts.

\end{abstract}
\section {Introduction}

It has been suspected for a long time that the large number
 of  extremely  bright
QSOs or radiosources may arise from magnification by foreground
gravitational structures (\cite{thomas95} and references therein).
Knowledge of the magnification
bias function provides constraints on the distribution of mass
condensations, the density parameter $\Omega$ of the Universe, as well
as the luminosity function of QSOs.
This has stimulated surveys to search for an excess of galaxies
close to the line of sight of the brightest QSOs; however inconclusive
results emerged.
For small angular separations, the detection is difficult without
large coronographs (blooming of the detector, instrumental scattering
of light).  The results are still controvertial but
nevertheless, on small scale ($< 5"$), an
excess has been observed which corresponds to an
overdensity of galaxies $q=1.4 \pm 0.5 $ for
 bright QSOs with M$_v<$-28 and z$>$1.5.
 The result seems to be in agreement with theoretical
predictions of magnification bias by individual
galaxy lenses (\cite{thomas95}).
Correlations on scales larger than one arcminute gave negative results
until \cite*{fugmann90} announced a correlation between the distribution
of 1-Jy distant radiosources (z $>$ 1) and the Lick
 catalogue of galaxies (z $<$ 0.1). This claim prompted the
further statistical analysis of \cite*{bartelmann93b},
who verified Fugmann's result with  the  sensitive
Spearman rank order correlation test.
They confirmed the existence of a correlation between the
foreground galaxies of both the Lick and IRAS catalogues and
radiosources with  z$>$1, on angular scales larger
than 10 arcminute (\cite{bartelmann93a}, \cite{bartelmann94a})!
The test does not give the strength
 of the correlation but we know that
 individual lensing galaxies cannot  produce correlations
on so large a scale (\cite{bartelmann93b}). A recent
photographic plate analysis (\cite{benitez95})  of the galaxy distribution
provides additional observational evidence that  the correlation is real.
\par
The lensing hypothesis would more easily explain the
correlation if the lensing agents have  larger cross sections than galaxies.
The most natural candidates are the numerous galaxy
clumps distributed in the Large Scale Structures of
the Universe (hereafter LSS) if a substantial
fraction of them have almost the critical surface mass density.
In fact, the  excess of QSOs and radiosources
around the Zwicky, the Abell and the ROSAT clusters
 reported recently (\cite{bartelmann94b}, \cite{Seitz95a})
 already supports the idea that cluster-like structures may
 play a significant role in magnifying a fraction of bright
quasars.
If this hypothesis is true
these massive, not  yet detected deflectors in visible
could show up through their weak lensing effects
on  the background galaxies.  \par

The gravitational weak lensing analysis has recently proved
to be a promising technique to
map the projected mass around
clusters of galaxies (\cite{kaiser93},
\cite{bonnet94}, \cite{fahlman94}, \cite{smail94}). Far from the
centers of such mass condensations, background galaxies are  weakly
stretched perpendicular to  the gradient of the
gravitational field. With the  high surface density of background galaxies  up
to $V=27.5$ ($\approx 43 $ faint sources per square arcminute with $V>25$)
the local shear (or polarization of the images) can be recovered from the
 measurement of the  image distortion of weakly lensed  background
galaxies averaged over a sky aperture with typical radius of 30
arcsec.
The implicit assumption that the magnification matrix is constant
on the scanning aperture is not always valid  and this observational
limitation  will be discussed later \par

The shear technique was also used  with success to detect
large unknown deflectors in front of the doubly imaged quasar Q2345+007
(\cite{bonnet93}).
This QSO pair has an abnormally high angular separation,
though no strong galaxy lens is visible in its neighbourhood.
The shear pattern revealed the
presence of a cluster mass offcentered  at one arcminute
north-east from the double quasar, which contributes to the
large angular separation. Further ultra-deep photometric observations in
the visible and the near infrared have {\it a posteriori} confirmed the
presence of the cluster centered on the center of the shear pattern and
detected a  small associated clump of galaxies as well, just on
the QSO line of sight. Both lensing agent are at a redshift
larger than 0.7 (\cite{mellier94b}, \cite{fisher94}, \cite{pello95}).
The predicting capability of the weak lensing was quite remarkable since it
{\it a priori} provided a better signature of the presence of
a distant cluster than the actual overdensity of galaxies, which in
the case of Q2345+007 was almost undectable
without a deep "multicolor" analysis.
\par
On a theoretical side, numerical simulations in standard adhesion HDM or CDM
models (\cite{bartelmann92}) can predict the occurrence of
quasar magnification. They have shown
that the large magnifications are correlated with the highest amplitudes
of the shear, which intuitively means that the largest
weak lensing magnifications are in the immediate
vicinity of dense mass condensations.
For serendipity fields they found from their simulations
that at least 6\% of background sources should have
a shear larger than 5\%. However, for a subsample of rather bright
radiosources or QSOs the probability should be larger, so that we can
reasonably expect quasar fields with a shear pattern above the detection
level.
\par
Since we can detect shear as faint as 3\% (\cite{bonnet94}), both
observational and theoretical arguments
convince us to start a survey of the presence of weak shear around
several  bright radiosources.
In practice, mapping the shear requires exceptional subarcsecond
seeing ($<$0.8 arcsec.) and long exposure times,
typically 4 hours in V with a four meter class telescope.
Observations of  a large unbiased selected sample of QSOs will
demand several years and before promoting the idea of a large survey
we decided to probe a few bright QSO fields
where  a magnification bias is more likely.

\begin{table*}[t]
\begin{tabular}{lccccccccc}
object &  $\alpha_{50}$  &  $\delta_{50}$  & $m_V$ & z & flux &
Tel./Instr. &  exp. & numb.  & seeing \\
 &    &   &  & &  &  &  time & files & (arcsec.) \\
\hline
PKS0135-247 & 01 \ 35 \ 17.17 & -24 \ 46 \ 11.7 & 16.98 & 0.832 & 1.70 &
NTT/SUSI &  16500 & 10 & 0.75 \\
PKS1508-05 & 15 \ 08 \ 14.94 & -05 \ 31 \ 49.0 & 17.20 & 1.191 & 2.43 &
NTT/SUSI &  13500 & 5 & 0.76 \\
PKS1741-03 & 17 \ 41 \ 20.62 & -03 \ 48 \ 49.0 & 19.00 & 1.057 & 3.65 &
NTT/SUSI &  23700 & 13 & 0.68 \\
3C446 & 22 \ 23 \ 11.08 & -05 \ 12 \ 17.8 & 18.03 & 1.404 & 3.87 &
NTT/SUSI &  19700 & 9 & 0.66 \\
Q1622+238 & 16 \ 22 \ 32.40 & +23 \ 52 \ 01.0 & 18.18 & 0.927 & -- &
CFHT/FOCAM  & 18000 & 10 & 0.78 \\
\hline
\end{tabular}
\caption{Observational data for the 5 QSOs fields. The V magnitude
stars. The radioflux is the 5009 MHz value from the
1Jy catalogue. The total exposure time corresponds to the coaddition
of several individual images with 30-45 minutes exposure time.
The seeing is the FWHM of stars on the composite image }
\end{table*}

In this paper, we report on a preliminary tests at CFHT and ESO of
five sources at $z \approx 1$.  The analysis of the shape
parameters and the shear is based on the \cite*{bonnet95} technical
paper, with some improvements to measure very weak ellipticities.
Due to instrumental difficulties
only one, Q1622+238, was observed at CFHT. Nevertheless, we found a
strong shear pattern in the immediate vicinity of the quasar quite similar
to the shear detected in the QSO lens
Q2345+007 (\cite{bonnet93}). The QSO is magnified
by a previously unknown distant cluster of galaxies.
The four other QSOs were observed with the imaging camera SUSI at the NTT
with a significantly lower instrumental distortion but
with a smaller field of view. In this case the limited
size of the camera makes the mapping of strong deflector like in Q1622+238
 harder. However, with the high image quality of SUSI it
is possible to see on the images a clear correlation between the amplitude and
direction of the shear and
the presence of foreground overdensities of galaxies.
Some of them are responsible for a magnification bias of the QSO.
\par
By comparing the preliminary observations at CFHT and ESO
we discuss  important observational issues, namely the
need for a perfect control of image quality and
a large field of view. We also show that invisible masses associated
 with groups and poor clusters of galaxies can be seen
through their weak lensing effect with NTT at ESO. These groups
of galaxies may explain the origin of a large angular correlation between
the distribution of distant radiosources ($z>1$) and
the distribution of low redshift galaxies ($z<0.3$)
The study of the correlation between the local shear and
nearby overdensity of foreground galaxies (masses) will be investigated
in following papers after new spectrophotometric
observations of the lensing groups.
\par
\section{Selection and observations of the sources}

The double magnification bias hypothesis maximises the probability of a
lensing effect for luminous distant sources (\cite{borgeest91}).
Therefore whenever possible we try to select
sources that are  both bright in radio (F$>$ 2Jy, V $<$18).
We also looked at quasars with absorption lines at lower redshift,
 to know if some intervening matter on the
lines of sight is present.  The QSOs are chosen at nearly
the mean redshift of the faint background galaxies (z from 0.8 to 1.)
used as an optical template to map the shear of foreground deflectors.
So far, we have observed 5 QSOs at redshift about 1 with a V magnitude
and radio flux in the range
from 17 to 19 and 1.7 to 3.85 respectively (Table 1).\par

Except Q1622+238 (z=0.97) which was suspected to have a faint
group of galaxies nearby (\cite{hintzen91}),
the 4 other candidates
(PKS0135-247, PKS1508-05, PKS1741-03, and 3C446.0)
have been only selected
from  the \cite*{hewitt87} , and the \cite*{veronveron85}
catalogues, choosing those objects with
 good visibility during the observing runs. The V magnitude of
each QSO was determined with an accuracy better than 0.05 mag. rms
from faint \cite*{landolt92} calibration stars (Table 1).\par

The observations started simultaneously in June 1994 at the ESO/NTT with
SUSI and at CFHT with FOCAM, both with excellent seeing conditions
($<$0.8") and stable transparency. For the second run at ESO in
November 1994, only one of the two nights
has good seeing conditions for the observation of PKS0135-247.
We used the 1024$\times$1024 TeK and the 2048$\times$2048 LORAL CCDs
with 15 micron pixel, which correspond to 0.13"/pixel at the
NTT and 0.205"/pixel at CFHT, and typical fields of view of 2' and
7' respectively.  In both cases we used a standard shift and add
observing technique with 30 to 45 min exposures.
The resulting field of view is given in table 2.
The total exposure was between 16500 and 23700 seconds in  V  (Table 1).
The focusing was carefully checked between each individual exposure.
After prereduction of the data with the IRAF software package, all
frames were  coadded leading to a composite image with an effective seeing of
0.78" at CFHT and 0.66"-0.78" at NTT
(Table 1). Although the seeing was good at CFHT
we are faced with a major difficulty
when trying to get a point spread function for stars (seeing disk) with
 small anisotropic deviations from circularity less than $b/a =0.05$ in
 every direction).
This limitation on the measurement of the weak shear amplitude
will be discussed more explicitly in the following section. \par

\begin{figure*}
\picplace{10cm}
\caption{Figure 1a: NTT Field of view of PKS1741 which was used as a star
template to study the instrumental distortion of the SUSI camera. Figure 2b:
plot of the apparent residual "shear amplitude" of the stars on 5 points of
the field where the galaxy shears are determined in other NTT images;
figures 4, 5, 6}
\end{figure*}

\section{Measurement of the shear}

\begin{figure}
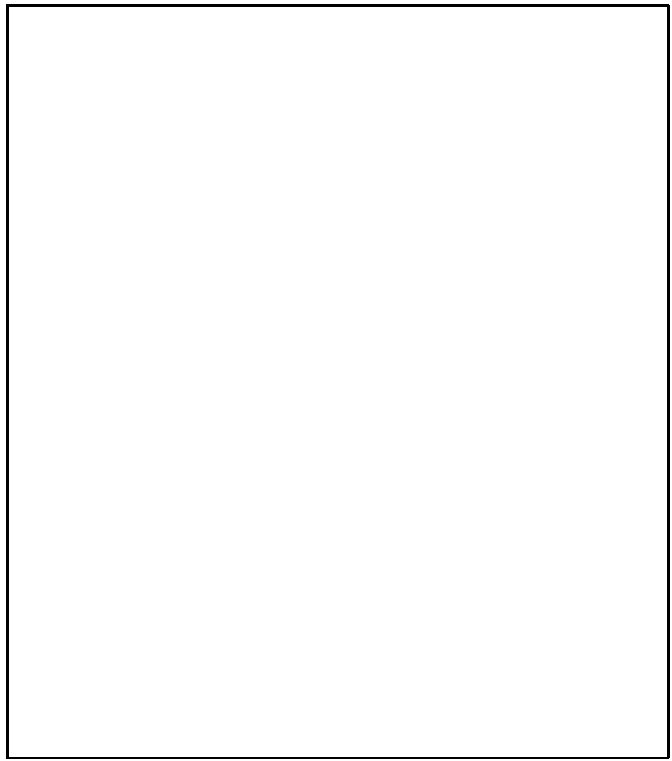

\picplace{10cm}
\caption{Histogram of the independent measurements of the
axis ratio $b/a$ in all the fields with a scanning aperture of
30 arcsec. radius.
The peak around 0.99 is representative of the noise level that
defines a
threshold of amplitude detection near 0.985.}
\end{figure}

The measurements of the shear patterns have been obtained from an average
of the centered second order momenta as computed by Bonnet and Mellier
(1995) of all individual galaxies in a square
aperture (scanning aperture size: 57 +3/-5 arcsec.)
containing at least 25 faint galaxies with V between 25 and
27.5 (Table 2). Because very elongated objects increase the dispersion
of the measurement of the averaged shape parameters (see Bonnet and
Mellier 1995, Fig. 4), and blended galaxies give wrong ellipticities,
we rejected these objects from the samples.
The direction of the polarization of background galaxies
is plotted on each QSO field (Figures 3b, 3d, 4b, 5b, 6b)
at the barycentre of the 25
background galaxies that are used to calculate the averaged shear. Each
plot has the same amplitude scale for comparison between
images and the instrumental distortion found from a star field analysis (Figure
1b).
This explains why the mapping is not rigorously made with a
regular step between each polarization vector on the figures.
The small step variation reflects
the inhomogeneity of the distribution of background sources.
For the exceptional shear pattern of Q1622+238,
a plot with a smaller sampling in boxes of
22 arcsec. gives a good view of the coherence of the shear
(Figure 3b).
All other maps are given with a one arcminute box, including figure 3d,
so that each measurement of the shear is completely independent.
For quantitative study the coordinates  of each measurement
are given on table 3
with the value of the apparent amplitude $1-b/a$ and the direction of the
shear.
 The ellipticity  $ e = 1 - b/a$
given in Table 3 is drawn on the various fields with the same scale.
\par
A description of the technique used to map the shear
can be found in  \cite*{bonnet95}. We have only improved
when necessary the method to  correct the instrumental distortion
in order to detect apparent shear on the CCD images
down to a level of about 2.0\% (Figure 2).
Notice that we call here "apparent shear" the observed shear
on the image which is not corrected for seeing effects and which is
averaged
within the scanning aperture. To achieve this goal we observed at NTT,
in similar conditions as other radiosources, the field
of PKS1741-03 which contains approximately
$26 \pm 6$ stars per square arcminute (Figure 1a,b).
After a mapping of the instrumental distortion of stars
we have seen that  prior to applying the original \cite*{bonnet95} method,
it is possible to restore an ideal circular seeing disk with a
gaussian distribution of energy for stars in the field
(pseudo deconvolution).
The correction almost gives  conservation of the seeing effective
radius with:
\begin{equation}
s = \sqrt{ <a^*> \times <b^*>} \ ,
\end{equation}
where $<a^*>$ and $<b^*>$  are the mean of the major and minor axes of
individual
stars in the field.
\par
With SUSI the original spread function measured
from stars is relatively stable over
the 2.2$\times$2.2 square arcmin. field and can be extraordinary good.
The blurring of images due to the rotator accuracy
when compensating for field rotation was found to be less than 0.004"/hour
at the edge of the field, and
the tracking errors are often nearly 0.04"/hour.\par
This level of instrumental distortion corresponds to a
mean ellipticity
of all stars in the field of $1-<b/a>$ = 0.015.
Therefore  Bonnet \& Mellier's method of correction
can generally be used directly without significant changes in the results.
After the correction, the systematic residual instrumental distortions
are often below
the intrinsic sensitivity of the method, which is limited by the
unknown distribution of source shapes within the resolution aperture
(Figure 1b).
However we verify with the  PKS1741-03 field that
the restoration  of the circularity of the spread function
can give a residual "polarization" of stars in the field as low as
$1-<b/a> = 0.0009 \pm 0.0048$ (dispersion).
\par
In fact the restoration of the point spread function
appeared to be more difficult with CFHT images
because of a higher level of instrumental distortion whose
origin is not yet completely determined: guiding errors, atmospheric
dispersion, larger mechanical flexure of a non-azimuthal telescope,
 3 Hz natural oscillation of the telescope (P. Couturier, private
communication),  optical caustic of the
parabolic mirror, and indeed greater difficulties in getting excellent
image quality on a larger field.  Thus, the level of instrumental
distortion measured on stars is currently $1-<b/a>$ = 0.08-0.12
with complex deviations from a circular shape.
After the restoration of an ideal seeing spread function
we are able to bring the shear accuracy of CFHT images
to a level of 0.03. But like the classical measurement of light polarization
it should be far better to start the observations with a level of
instrumental polarization as low as possible.
\par
In summary we are now able to reach the intrinsic limitation
of  Bonnet \& Mellier's method on the measurement
of the shear amplitude at NTT with a typical resolution
of about 60 arcsec. diameter (25-30 faint galaxies per resolution element)
with a rms error of about 0.015 (Figure 2).
Below this value the determination of the amplitude of the shear
is meaningless although the direction may
still be valid.
At CFHT the detectivity is almost two times less but the field is larger.
We are currently developing methods to
correct the instrumental distortion at the same level we get with the NTT.
This effort is necessary for future programmes with the VLT
which would be aimed toward the mapping of Large Scale Structures
(shear of 0.01) with a lower spatial resolution ($>10$ arcminute apertures).
\par
\section{Results}

In this section we discuss the significance of the shear
pattern
in each QSO field and the eventual correlation with the isopleth
or isodensity curves  of background galaxies with
$20 < V < 24.5$. For a fair comparison both the isopleth
(surface density numbers) or
isoluminosity curves (isopleth weighted by individual luminosity) are
smoothed with a gaussian filter having nearly
the resolution of the shear map (40" FWHM).
\par
\begin{enumerate}
\vskip 3truemm \item {\em Q1622+238}
\par
A coherent and nearly elliptical shear pattern is
detected
with an apparent amplitude  0.025$\pm$ 0.015 at a distance ranging from
50" to 105" of the QSO (Figure 3b). The center of the shear
can be calculated with
the centering algorithm described by Bonnet \&
Mellier. The inner ellipses  in figure 3b  show the position of the
center at  the 1, 2 and 3 $\sigma$  confidence level. It
coincides with a cluster of galaxies
identified on the deep V image 10 arcsec South-East from the
QSO
(Figure 3c). The external contour of the isopleth map in figures 3c
corresponds to a density excess of galaxies of twice the averaged
values on the field for  a 30 arcsec circular aperture. The
isoluminosity map shows a light concentration even more compact than
the number density map. About
70\% of the galaxies of the  condensation have a narrow magnitude range
between $V= 24$ and $24.5$ and are concentrated around a bright
galaxy with $V=21.22 \pm  0.02$. This is typical for a
cluster of galaxies.
A short exposure in the I band gives a corresponding magnitude
$I=19.3 \pm 0.1$ for the bright central galaxy.
A simple use of the magnitude-redshift relationship from a Hubble
diagramme and the $(V-I)$ colors of the galaxy
suggest a redshift larger than 0.5. By assuming such a redshift

\begin{table}
\begin{tabular}{lccc}
Object & field  & Ng/N$_G$ & Mag  \\
       & (pixels)/(arcsec.) &  & range  \\
\hline
PKS0135-247 & 930$\times$970 / 121$\times$126 & 112/48 & 22-25  \\
PKS1508-05 & 951$\times$947/124$\times$123 & 152/33 & 22-24.5  \\
PKS1741-03 & 888$\times$818/115$\times$106 & 99/27 & 22-24.5  \\
3C446 & 925$\times$946/120$\times$123 & 102/65 & 22-25  \\
Q1622+238 & 1871$\times$1855/385$\times$382 & 1652/486 & 21-24.5  \\
\hline
\end{tabular}
\caption{Table 2: Number Ng of (background) galaxies from V=22 to 24.5
which are
used to trace isopleth and number N$_G$ of (distant) galaxies from V=25 to
27.5.
detected on each observed field}
\end{table}

it is possible to mimic the shear map with
a deflector velocity dispersion of at least 500 km/sec.
After a correction for the seeing effect with the \cite*{bonnet95} diagram
 and taking into account the local shear of the lens at the exact
location of the QSO we can estimate that the magnification bias
could be exceptionally high in this case
($>$ 0.75 magnitude). Further spectrophotometric observations of the field
are needed to get a better description of the lens. It is even possible
that multiply imaged galaxies are present at the center of this newly
discovered cluster.

\begin{figure*}
\picplace{10cm}
\caption{ Figure 3a: CFHT field of view of Q1622+238 in V. North is at the top.
Figure 3b: Shear map of  Q1622+238 with a resolution step of 22 arcsec. The
ellipses shows the position of the center of the central shear with the
1, 2, 3 $\sigma$ confidence level. The center almost coincides with
a distant cluster clearly visible on figure 3c.}
\end{figure*}

\begin{figure*}
\picplace{10cm}
\caption{ Figure 3c: Zoom at the center of the field of view of Q1622+238.
The distant cluster around the bright central elliptical galaxy E is
clearly identified on this very deep V image.
Figure 3d: Shear map of  Q1622+238 with a resolution step of 60 arcsec.
similar to the resolution on other NTT fields. The
ellipses shows the position of the center of the central shear with the
1, 2, 3 $\sigma$ confidence level. The center almost coincides with
a distant cluster clearly visible on figure 3c.}
\end{figure*}
\par
\vskip 3truemm \item {\em PKS1741-03}

This first NTT field was chosen for a dedicated study of the instrumental
distortion of the SUSI instrument.
Indeed it is crowded with stars and the mapping of the
isopleth was not done due to large areas of the sky occulted
by bright stars.
\par
The center of the field of PKS1741-03 shows a faint compact group of
galaxies (marked g on fig 1a).
A detailed investigation of the alignment of individual faint galaxies
nearby shows that a few have almost orthoradial
orientation to the center of the
group.  The amplitude of the "apparent" shear on the fig 1b is low
probably because it rotates within the scanning aperture around a
deflector having an equivalent velocity dispersion lower than 400 km/s.
Outside  the box the apparent shear is already below the $1-<b/a> = 0.015$
threshold level and it is not possible to detect the circular
shear at distance from the group larger than one arcmin.  This remark is
important because it illustrates the limitation of the method in detecting
lenses with a  $1-<b/a> = 0.015$ on angular scales smaller than the scanning
aperture. Therefore a low amplitude of the shear
on the scanning aperture could be the actual signature of a small deflector
rather than a sky area with a low shear!
Although the compact group is only 30 arcsec South-East of the QSO
it might contribute to a weak lensing of  PKS1741-03
but it is difficult to get a rough estimate of
the amplitude of the magnification bias.

\begin{figure*}
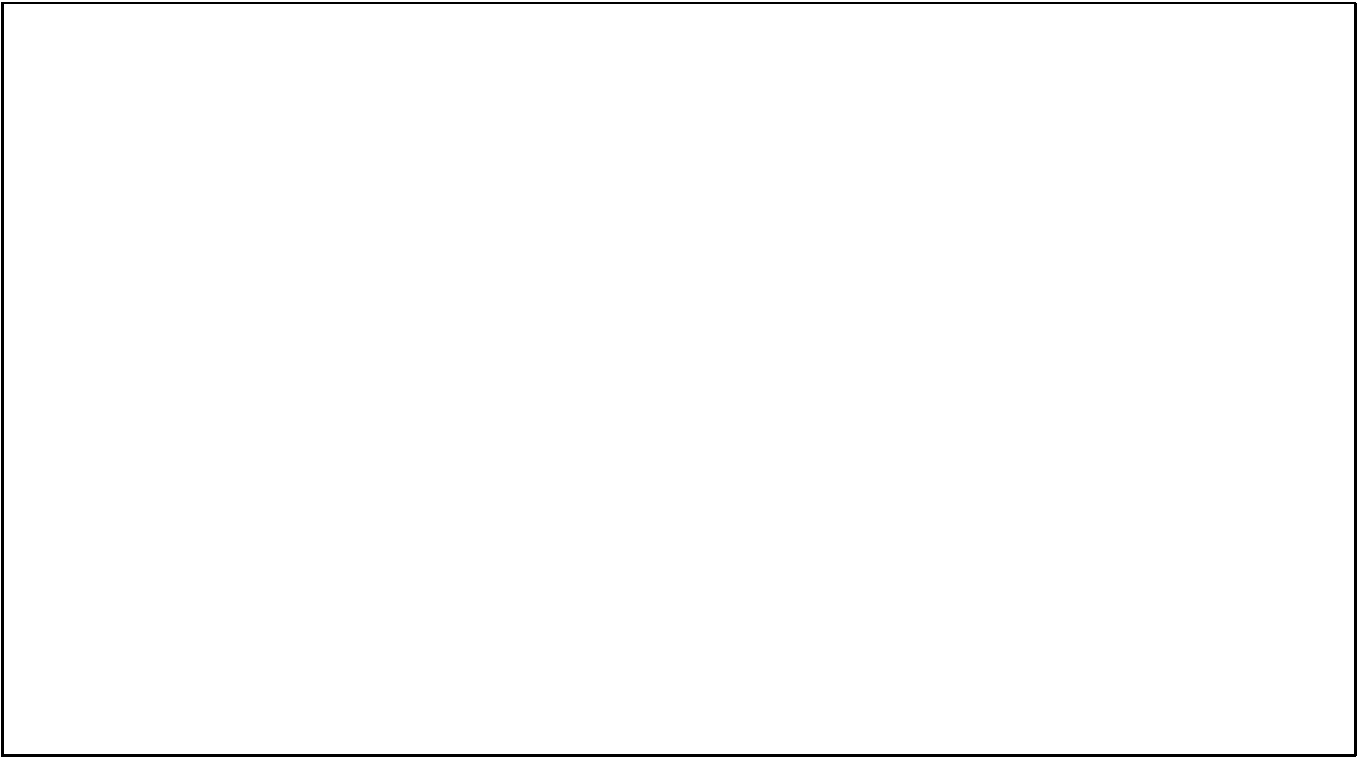

\picplace{10cm}
\caption{Figure 4a: NTT field of view for PK0135. North is at the top. Note
the group of galaxies around g1, g2, g3 and g4 responsible for a coherent
shear visible on figure 4b}
\end{figure*}

\begin{figure*}
\caption{Figure 5a: NTT field of view for PKS1508. Note
the North-West group of galaxies near the brighter elliptical E
responsible for a larger amplitude of the shear on figure 5b and the
small clump of galaxies g right on the line of sight of the QSO.}
\end{figure*}
\par
\vskip 3truemm \item {\em PKS1508-05}

This is the second bright radiosource of  the sample.
At one arcminute North-West there is also a group around a bright galaxy ($G$)
which could be responsible for a large shear. This distant group or cluster
may contribute to a weak magnification by itself, but there is
also a small clump of galaxies in the close vicinity
of the radiosource with the brightest member at a distance of 8 arcseconds
only.
The situation is similar to the case of the multiple QSO 2345+007
(\cite{bonnet93}).
This could be the dominant lensing agent
which provides a larger magnification bias, especially
if the nearby cluster has already provided a substantial part
of the critical projected mass density.
\par

\vskip 3truemm \item {\em 3C446}

The radiosource is  among the faintest in the optical (table 1).
There is a loose group of galaxies at 40 arcsec South-West from the
QSO. The orientation of the shear with respect to the group
of galaxies  can be reproduced with a rough 2D
simulation (\cite{hueppe95}) although at first look
it was not  so convincing as the PKS0135-247 case.
The lensing configuration could be similar to
PKS1508-05 with a secondary lensing agent G near the QSO (fig 6a,b).
Surprisingly there is also a large shear amplitude which is not apparently
linked to an overdensity of galaxies in V in the North-East corner.
In such a case it is important to confirm
the result with an I image to detect possible distant groups
at a redshift between 0.5 and 0.7.
{\it A contrario} it is important to mention that
the shear is almost null in the North-West area of
the field which actually has no galaxy excess visible
in V  (fig 6b).
\par

\begin{figure*}
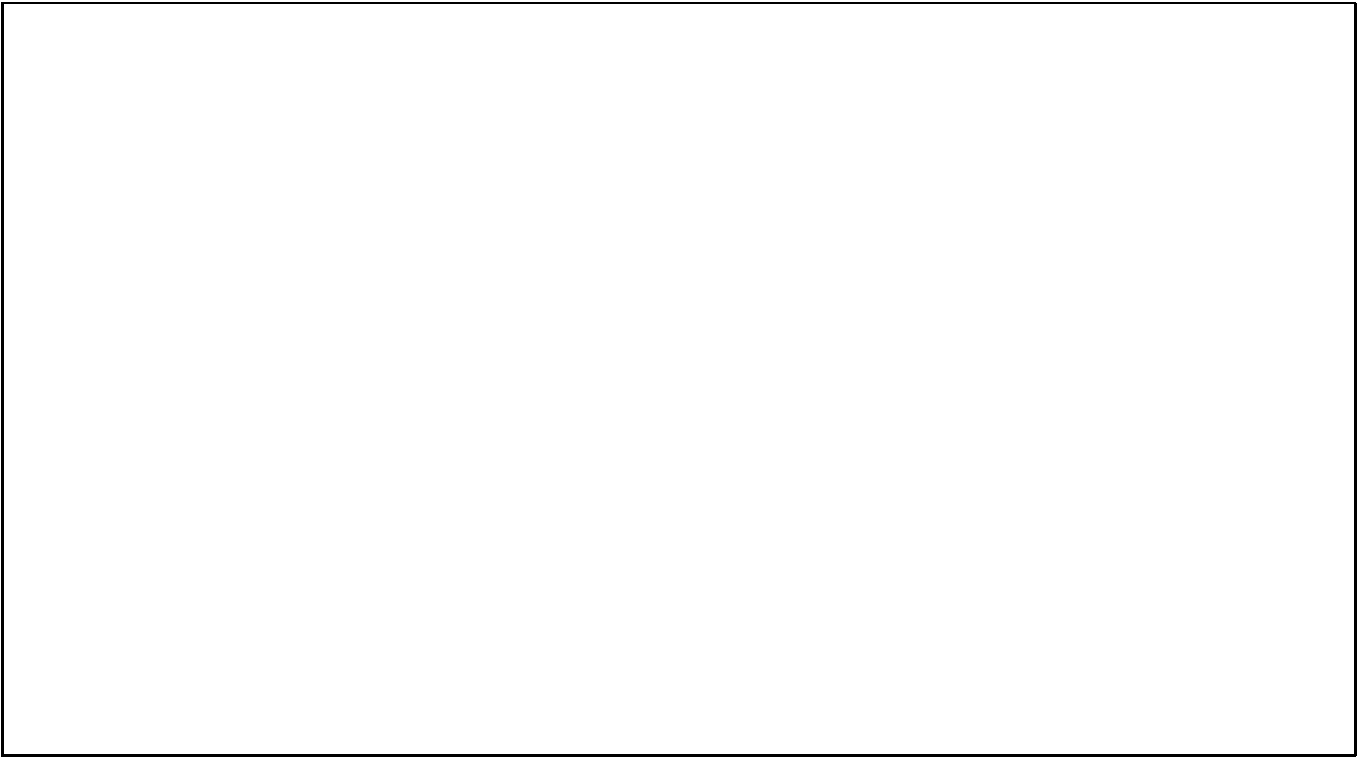

\picplace{10cm}
\caption{Figures 6a: NTT field of 3C446. Note on figure 6b
the shear pattern relatively
to the isopleth of possible foreground groups and
the galaxies g on the line of sight of the QSO}
\end{figure*}

\end{enumerate}

\section{Discussion}

Due to observational limitations on the visibility of radiosources
during the observations the selection criteria were actually
very loose as compared with what we have proposed in Section 2
for a large survey.  The results we present here must be considered as
a sub-sample of QSOs with a moderate possible bias.
Nevertheless, for at least 3 of the sources there are some lensing agents
 which are associated with foreground groups or clusters of galaxies
that are detected and correlated with  the shear field.
 For the 2 other cases the signature of
a lensing effect is not clear but cannot be discarded from the
measurements. All the radiosources may have a magnification bias
enhanced by a smaller clump on the line of sight or even an
(unseen) foreground galaxy lying a few arcsec from the radiosource
(compound lens similar to PKS1508).
The occurrence of coherent shear associated with groups in the field
of the radiosources is surprisingly high. This might mean
that a lot of groups or poor clusters which are not yet identified
contain a  substantial part of
the hidden mass of LSS of the Universe below $z= 0.8$.
Some of them responsible for the observed apparent shear
may be the most massive progenitor clumps of rich clusters still
undergoing merging.
\par
Although these qualitative results already
represent a fair amount of observing time we are now quite
convinced that all of these fields should be reobserved, in
particular in the I and K bands, to assess the nature of the deflectors.
Spectroscopic  observation of the brightest members
of each clump is also necessary to determine
the redshift of  the putative deflectors.
This is an indispensable step to  connect the shear pattern to a
quantitative amount of lensing mass and to link the polarization map
 with some dynamical parameters of visible matter, such as  the velocity
dispersion for each deflector, or possibly the X-ray emissivity.
at the present time,  we are only able to say that
there is a tendency for a correlation between the shear and
light overdensity (\cite{fort94}).
\par
\begin{table}
\begin{tabular}{lcccc}
x & y & angle & 1-b/a   \\
  &  & (degrees) & \\
\hline
PKS0135-247 & & &  \\
274 & 202 & -83 & 0.012 \\
258 & 741 &  0  & 0.027 \\
698 & 228 &  67 & 0.006 \\
713 & 681 & 42  & 0.026\\
473 & 408 & 69  & 0.021 \\
 & & & \\
PKS1508-05 & & &  \\
270 & 272 & 73 & 0.012 \\
216 & 712 & -62 & 0.009 \\
704 & 251 & -65 & 0.026 \\
745 & 695 & -51 & 0.032 \\
453 & 463 & -85 & 0.016 \\
 & & & \\
PKS1741-03 & & &  \\
255 & 268 & 12 & 0.013 \\
256 & 613 & -56 & 0.013 \\
694 & 286 & -63 & 0.006 \\
716 & 620 & -80 & 0.017 \\
473 & 425 & 37 & 0.003 \\
 & & & \\
3C446 & & & \\
247 & 261 & 86 & 0.014 \\
284 & 707 & 74 & 0.027 \\
688 & 271 & 66 & 0.015 \\
720 & 693 & 70 & 0.002 \\
452 & 421 & -69 & 0.021 \\
 & & & \\
Q1622+238 & & & \\
209 & 187 & -54 & 0.005 \\
201 & 583 & 16 & 0.006 \\
203 & 978 & -50 & 0.008 \\
196 & 1399 & -28 & 0.013 \\
259 & 1665 & -46 & 0.009 \\
629 & 201 & -50 & 0.019 \\
588 & 596 & -65 & 0.033 \\
610 & 996 & 53 & 0.014 \\
603 & 1398 & 4 & 0.022 \\
616 & 1682 & 4 & 0.006 \\
984 & 194 & 24 & 0.013 \\
999 & 590 & 14 & 0.015 \\
996 & 979 & 65 & 0.010 \\
992 & 1402 & 20 & 0.013 \\
983 & 1718 & -20 & 0.025 \\
1407 & 202 & -43 & 0.013 \\
1386 & 598 & 78 & 0.051 \\
1403 & 999 & 13 & 0.002 \\
1407 & 1401 & 42 & 0.013 \\
1420 & 1708 & -69 & 0.012 \\
1698 & 210 & 52 & 0.015 \\
1715 & 595 & 44 & 0.022 \\
1723 & 1027 & 33 & 0.025 \\
1716 & 1387 & 41 & 0.014 \\
1650 & 1676 & -49 & 0.006 \\
 & & & \\
\hline
\end{tabular}
\caption{Summary of relevant data for weak lensing measurements.
X and y are the averaged center positions (in pixel) of each independent
box shown in the previous figures. The orientations and the axis ratios
also correspond to the values associated with each individual
line of the figures.
}
\end{table}

\par
{}From the modelling point of view,  simulations have been done
and reproduce fairly well the direction of the shear pattern with a
distribution of mass that follows most of the light distribution
given by the isopleth or isoluminosity contour of the groups
in the fields.
Some of these condensations do not play any role at all and are probably
too distant to deflect the light beams. Unfortunately, in order  to make
accurate  modelling it is necessary to have a good
estimate of the seeing effect on the amplitude of the shear by
comparing with HST reference fields, and good redshift determinations as
well of the possible lenses to get their gravitational weight in the field.
It is also important to consider more carefully the effect of
convolution of the actual local shear which varies at smaller scales
than the size of the scanning beam (presently about one arcminute size). This
work is now being done  but is also waiting for
more observational data to actually start to study the gravitational mass
distribution of groups and poor clusters of galaxies in the field of
radiosources.
\par
\section{Conclusion}

The shear patterns observed in the fields of five bright QSOs,
and the previous detection of a cluster shear
in Q2345+007  (\cite{bonnet93})
provide strong arguments in favor of the \cite*{bartelmann93b} hypothesis
to explain the large scale correlation
between radiosources and foreground galaxies. The LSS could be
strongly structured by  numerous
condensations of masses associated with groups of galaxies. These groups
produce significant weak lensing effects that can be
detected. A rough estimate of the magnification bias is given by
the polarization maps around these radiosources. It could
sometimes be higher than half a magnitude and even much more
with the help of an individual galaxy deflector at a few arcsec.
of the QSO line of sight.
The results  we report here also
show that we can study with the weak shear analysis
the distribution of density peaks of (dark) massive gravitational structures
(ie $\sigma > $ 500 km/s)  and characterise their association with
overdensities of galaxies at moderate redshift (z from 0.2 to 0.7).
\par
A complete survey of a large sample of radiosource fields will have
strong cosmological interest for the two aspects we mentioned above.
Furthermore, the method can be used to probe the
intervening masses which are associated with the absorption lines in QSOs
or to explain the unusually high luminosity of distant sources like
the ultraluminous sources IR 10214+24526 (\cite{serjeant95}) or the most
distant radio-galaxy 8C1435+635 (z=4.25; \cite{lacy94}).
\par
Therefore we plead for the continuation of systematic measurements
of the shear around a sample of bright
radiosources randomly selected with the double magnification
bias procedure (\cite{borgeest91}).
Our very first attempt encountered some unexpected obstacles related
to the limited field of view of CCDs or the correction of
instrumental distortion. It seems
that they can be overcome in the near future. We have good hopes that
smooth distributions of mass associated with
larger scale structures like filaments and wall structures
could be observed
with a dedicated wide field instrument that minimizes all instrumental
and observational systematics, or still better with
a Lunar Transit telescope (\cite{fort95}).

\paragraph{Acknowledgements}
We thank P. Schneider, N. Kaiser, R. Ellis, G. Monet, S. D'Odorico, J.
Bergeron and P. Couturier for their enthusiastic support and
for useful discussions for the preparation of the observations. The data
obtained at ESO with the NTT would probably not have been so excellent
without the particular care of
P. Gitton for the control of the image
quality with the active mirror. We also thank P. Gitton
for his helpful comments and T. Brigdes for a careful reading
of the manuscript and the english corrections.
This work was supported by grants from the GdR Cosmologie and from the
European Community (Human Capital and Mobility ERBCHRXCT920001).

\bibliographystyle{alpha}
\bibliography{bib/biblio}

\end{document}